\documentclass[aoas,nameyear,seceqn,dvips]{arximspdf}
\usepackage{graphics}
\doi{10.1214/10-AOAS327}
\volume{4}
\issue{3}
\pubyear{2010}
\firstpage{1579}
\lastpage{1601}

\makeatletter
\newproclaim{remark}{Remark}
\newtheorem{theorem}{Theorem}[section]
\newproclaim{algorithm}{Algorithm}
\newcommand{\eqref}[1]{(\ref{#1})}
\makeatother

\begin{document}
\begin{frontmatter}

\title{Sparse logistic principal components analysis\\ for binary data}
\runtitle{Sparse logistic PCA}

\begin{aug}
\author[a]{\fnms{Seokho} \snm{Lee}\ead[label=e1]{lees@ufs.ac.kr}},
\author[b]{\fnms{Jianhua Z.} \snm{Huang}\thanksref{t1}\ead[label=e2]{jianhua@stat.tamu.edu}\corref{}}
\and
\author[c]{\fnms{Jianhua} \snm{Hu}\thanksref{t2}\ead[label=e3]{jhu@mdanderson.org}}

\thankstext{t1}{Supported in part by Grants from the National Science
Foundation
(DMS-06-06580, DMS-09-07170), the National Cancer Institute (CA57030),
the Virtual Center for Collaboration between Statisticians in the US
and China,
and King Abdullah University of Science and Technology (KAUST, Award
KUS-CI-016-04).}
\thankstext{t2}{Supported in part by Grants from the National Science
Foundation
(DMS-07-06818) and the National Institute of Health (R01-RGM080503A,
R21-CA129671).}
\runauthor{S. Lee, J. Z. Huang and J. Hu}

\affiliation{Harvard School of Public Health,
Texas A\&M University\\ and University of Texas M. D. Anderson Cancer Center}

\address[a]{S. Lee\\
Department of Statistics\\
Hankuk University of Foreign Studies\\
Yongin-si, Gyeonggi-do\\
449-771 Korea\\
\printead{e1}}

\address[b]{J. Z. Huang\\
Department of Statistics\\
Texas A\&M University\\
College Station, Texas 77843-3143\\
USA\\ \printead{e2}}

\address[c]{J. Hu\\
Department of Biostatistics\\
Division of Quantitative Sciences\\
University of Texas M. D. Anderson Cancer Center\\
Houston, Texas 77030-4009\\ USA\\
\printead{e3}
}
\end{aug}

\received{\smonth{1} \syear{2009}}
\revised{\smonth{1} \syear{2010}}

%
\begin{abstract}
We develop a new principal components analysis (PCA) type dimension
reduction method for binary data. Different from the standard PCA
which is defined on the observed data, the proposed PCA is defined
on the logit transform of the success probabilities of the binary
observations. Sparsity is introduced to the principal component
(PC) loading vectors for enhanced interpretability and more stable
extraction of the principal components. Our sparse PCA is formulated as
solving an optimization problem with a criterion function motivated
from a penalized Bernoulli likelihood. A Majorization--Minimization
algorithm is developed to efficiently solve the optimization problem.
The effectiveness of the proposed sparse logistic PCA method is
illustrated by application to a single nucleotide polymorphism data set
and a simulation study.
\end{abstract}

%
\begin{keyword}
\kwd{Binary data}
\kwd{dimension reduction}
\kwd{MM algorithm}
\kwd{LASSO}
\kwd{PCA}
\kwd{regularization}
\kwd{sparsity}.
\end{keyword}

\end{frontmatter}
%

\section{Introduction}\label{sec:intro}

Principal components analysis (PCA) is a widely used method for
dimensionality reduction, feature extraction and visualization of
multivariate data. Several sparse PCA methods have recently been
introduced to improve the standard PCA [e.g., Jolliffe, Trendafilov and Uddine (\citeyear{JTU03});
Zou, Hastie and Tibshirani (\citeyear{ZHT06}); Shen and Huang (\citeyear{SH08})]. By requiring the principal
component loading vectors to be sparse, sparse PCA methods yield PCs
that are more easily interpretable. Sparsity also regularizes the
extraction of PCs and thus makes the extraction more stable. Such
stability is much desired when the dimension is high, especially
in the so-called high-dimension low-sample-size settings. As extensions
of the standard PCA, however, these sparse PCA methods are mostly
suitable to variables of a continuous type, they are not generally
appropriate for other data types such as binary data or counts. Although
the basic objective of PCA, or its sparse version, can be achieved
regardless of the nature of the original variable, it is true that
variances and covariances have especial relevance for multivariate
Gaussian variables, and that linear functions of binary variables are
less readily interpretable than linear functions of continuous variables
[Jolliffe (\citeyear{J02})]. The goal of this paper is to develop a sparse PCA
method for binary data.

There are two commonly used definitions of PCA that give rise to the
same result.
PCA can be defined by finding the orthogonal projection
of the data onto a low dimensional linear subspace such that the
variance of the
projected data is maximized [Hotelling (\citeyear{H33})].
Alternatively, PCA can also be defined by finding the linear projection
that minimizes the mean squared distance between the data points and
their projections [Pearson (\citeyear{P01})]. Shen and Huang (\citeyear{SH08}) developed
their sparse
PCA method following the viewpoint of Pearson. Suppose
$\mathbf{y}_1,\ldots,\mathbf{y}_n\in\mathbb{R}^d$ are the $n$ data
points and
consider a
$k$-dimensional ($k<d$) linear manifold spanned by a
bases $\tilde{\mathbf{b}}_1,\ldots,\tilde{\mathbf{b}}_k$ with a
shift vector $\bolds{\mu}$.
According to Pearson, the PCA minimizes the following reconstruction error,
%
%
\begin{equation}\label{eq:rec-err}
\sum_{i=1}^n\Vert\mathbf{y}_i-(\bolds{\mu}+a_{i1}\tilde{\mathbf
{b}}_1+\cdots+a_{ik}\tilde{\mathbf{b}}_k)\Vert^2:
\end{equation}
subject to the constraint that $\mathbf{A}= (a_{ij})$ has orthonormal columns.
Usually the variables presented in $\mathbf{y}_i$ are scaled so that
they have the
same order of magnitude. Note that \eqref{eq:rec-err} is
a least squares regression if $a_{ik}$'s were known. In light of this
connection to regression and borrowing the idea from LASSO [Tibshirani (\citeyear{TIB96})], Shen
and Huang (\citeyear{SH08}) proposed to add an $L_1$ penalty $\|\tilde\mathbf
{b}_1\|_1 +
\cdots+\|\tilde\mathbf{b}_k\|_1$\vspace*{1pt} to the reconstruction error \eqref
{eq:rec-err} to obtain
sparse loading vectors $\tilde\mathbf{b}_1, \ldots, \tilde\mathbf
{b}_k$. Since the
reconstruction error \eqref{eq:rec-err} can be viewed as the negative log
likelihood up to a constant for the Gaussian
distributions with mean vectors
$\bolds{\theta}_i =\bolds{\mu}+a_{i1}\tilde{\mathbf{b}}_1+\cdots
+a_{ik}\tilde{\mathbf{b}}_k$ for
$i=1,\ldots,n$ and identity covariance, the method of Shen and Huang
can be
interpreted as a penalized likelihood approach for the sparse PCA. The
key idea of
the current paper is to replace the Gaussian likelihood by the Bernoulli
likelihood where $\bolds{\theta}_i$ will be the logit transform of
the success
probabilities. We refer to the proposed PCA method as sparse logistic PCA.
The relationship of the proposed sparse logistic PCA to the sparse PCA of
Shen and Huang is analogous to the relationship between logistic
and linear LASSO regression.

We develop an iterative weighted least squares algorithm to perform the
proposed sparse logistic PCA. Since the log Bernoulli likelihood is not
quadratic and the $L_1$ penalty function is nondifferentiable, the
optimization problem defining the sparse logistic PCA is not straightforward
to solve. Our algorithm applies the general idea of optimization
transfer or
Majorization--Minimization (MM) algorithm [Lange, Hunter and Yang (\citeyear{LHY00a}); Hunter and
Lange (\citeyear{HL04})]. By iteratively replacing the complex objective function with
suitably defined quadratic surrogates, each step of our algorithm solves
a weighted least squares problem and has closed form. The algorithm is easy
to implement and guaranteed at each iteration to improve the penalized PCA
log-likelihood. We show that the same MM algorithm is applicable when
there are missing data. We also develop a method for choosing the penalty
parameters and for choosing the number of important principal components.
PCA of binary data using Bernoulli likelihood has previously been studied
by Collins, Dasgupta and Schapire (\citeyear{CDS01}), Schein, Saul and Ungar (\citeyear{SSU03}) and de Leeuw (\citeyear{DeL06}),
but none of these works considered sparse loading vectors. As we demonstrate
using simulation and real data, sparsity can enhance interpretation of results
and improve the stability and accuracy of the extracted principal components.

Other approaches of sparse PCA are not as easily extendible to binary data.
Jolliffe, Trendafilov and Uddine (\citeyear{JTU03}) modified the defining maximum variance problem
of the
standard PCA by applying an $L_1$-norm constraint on the PC loading vectors
to obtain PCA with sparse loadings. Its use of sample variance makes it
unappealing for binary data. Zou, Hastie and Tibshirani (\citeyear{ZHT06}) rewrote PCA as a
regression-type optimization problem and then applied the LASSO penalty
[Tibshirani (\citeyear{TIB96})] to obtain sparse loadings. However, since the data appear
both as regressors and responses in their regression-type problem, the
connection of their approach to the penalized likelihood is not as
natural as
Shen and Huang (\citeyear{SH08}).

The rest of this article is organized as follows. In Section \ref{sec2} we
introduce the
optimization problem that yields the sparse logistic PCA and provides
methods for tuning parameter selection. Section \ref{sec:snps} applies
the sparse logistic PCA to a single nucleotide polymorphism data set and
compares it with the nonsparse version of logistic PCA. Section \ref{sec:MM} presents
a Majorization--Minimization algorithm for efficient computation of the
sparse logistic PCA and Section \ref{sec:missing} discusses how to handle missing data.
Results of a simulation study are given in Section \ref{sec:simu}. Section \ref{sec7} concludes
the paper with some discussion. The \hyperref[app]{Appendix} contains proofs of theorems.

\section{Sparse logistic PCA with penalized likelihood}\label{sec2}
\subsection{Penalized Bernoulli likelihood}\label{sec:model}

Consider the $n\times d$ binary data matrix $\mathbf{Y}=(y_{ij})$,
each row of
which represents a vector of observations from binary variables.
We assume that entries of $\mathbf{Y}$ are realizations of mutually independent
random variables and that $y_{ij}$ follows the Bernoulli distribution
with success probability $\pi_{ij}$. Let
$\theta_{ij}=\log\{\pi_{ij}/(1-\pi_{ij})\}$ be the logit
transformation of
$\pi_{ij}$. Define the inverse logit transformation
$\pi(\theta)=\{1+\exp(-\theta)\}^{-1}$. Then the success probabilities
can be represented using the canonical parameters as
$\pi_{ij}=\pi(\theta_{ij})$. The individual data generating
probability becomes
\[
\Pr(Y_{ij}=y_{ij})  =  \pi(\theta_{ij})^{y_{ij}}\{1-\pi(\theta
_{ij})\}
^{1-y_{ij}}
 =  \pi(q_{ij}\theta_{ij}),
\]
with $q_{ij}=2y_{ij}-1$ since $\pi(-\theta)=1-\pi(\theta)$. This
representation leads to the compact form of the log likelihood
as
%
%
\begin{equation}\label{eq:lik}
\ell =  \sum_{i=1}^n\sum_{j=1}^d \log\pi(q_{ij}\theta_{ij}).
\end{equation}
Note that the Bernoulli distributions are in the exponential family and
$\theta_{ij}$ are the corresponding canonical parameters.

To build a probabilistic model for principal components analysis of
binary data,
the $d$-dimensional canonical parameter vectors
$\bolds{\theta}_i=(\theta_{i1},\ldots,\theta_{id})^T$ are
constrained to reside
in a low dimensional manifold of $\mathbb{R}^d$ with the
dimensionality~$k$.
(The choice of $k$ will be discussed later in Section \ref{sec:k}.)
Specifically,
we assume that, for
some vectors $\bolds{\mu}$, $\tilde{\mathbf{b}}_1,\ldots,\tilde
{\mathbf{b}}_k \in\mathbb{R}^d$,\vspace*{1pt}
the vector of canonical parameters satisfies
$\bolds{\theta}_i = \bolds{\mu}+ a_{i1}\tilde{\mathbf{b}}_1 +
\cdots+ a_{ik}\tilde{\mathbf{b}}_k$
for $i =1, \dots, n$. We call
$\tilde{\mathbf{b}}_1,\ldots,\tilde{\mathbf{b}}_k$ the principal
component loading vectors
and the coefficients $\mathbf{a}_i=(a_{i1},\ldots, a_{ik})^T$ the
principal component
scores (PC scores) for the $i$th observation. Geometrically, the
vectors of canonical
parameters $\bolds{\theta}_i$ are projected onto the $k$-dimensional
manifold which
is the affine subspace spanned by $k$ PC loading vectors and translated
by the
intercept vector $\bolds{\mu}$. In matrix form, the canonical
parameter matrix
$\bolds{\Theta}=(\theta_{ij}) =(\bolds{\theta}_1,\ldots,\bolds
{\theta}_n)^T$ is represented as
%
%
\begin{equation}\label{eq:model}
\bolds{\Theta} =  \mathbf{1}_n\otimes\bolds{\mu}^T + \mathbf
{A}\mathbf{B}^T,
\end{equation}
where $\mathbf{A}=(\mathbf{a}_1,\ldots,\mathbf{a}_n)^T$ is the
$n\times k$ principal component
score matrix and $\mathbf{B}=(\tilde{\mathbf{b}}_1,\ldots,\tilde
{\mathbf{b}}_k)$ is the
$p\times k$
principal component loading matrix.
For identifiability purpose, we require that $\mathbf{A}$ has
orthonormal columns.

We target a method that can produce a sparse loading matrix, a loading
matrix with many zero elements. A sparse loading matrix implies variable
selection in principal components analysis, since each principal component
only involves those variables corresponding to the nonzero elements of
the loading vector.
We propose to perform variable selection using the penalized likelihood
with a sparsity inducing penalty.
Let $\mathbf{b}_j^T$ denote the $j$th row of $\mathbf{B}$. Then
\eqref{eq:model} implies that $\theta_{ij}=\mu_j+\mathbf
{a}_i^T\mathbf{b}_j$ where
$\mu_j$
is the $j$th element of $\bolds{\mu}$. The log likelihood
can be written as
%
%
\begin{equation}\label{eq:logL}
\ell(\bolds{\mu}, \mathbf{A}, \mathbf{B})  =  \sum_{j=1}^d \sum
_{i=1}^n\log\pi\{q_{ij} (\mu
_j+\mathbf{a}_i^T\mathbf{b}_j)\}.
\end{equation}
If $\mathbf{a}_i$ were observable, \eqref{eq:logL} is the log
likelihood for $d$
logistic regressions
\[
\operatorname{logit} P(Y_{ij}=1) = \mu_j + \mathbf{a}_i^T \mathbf{b}_j.
\]
This connection with logistic regression suggests use of the $L_1$ penalty
to get a sparse loading matrix, as in the LASSO regression [Tibshirani (\citeyear{TIB96})].

Specifically, consider the penalty
%
%
\begin{equation}\label{eq:pen}
P_{\bolds{\lambda}}(\mathbf{B})  =  \sum_{l=1}^k\lambda_l\|\tilde
{\mathbf{b}}_l\|_1
 =  \lambda_1\sum_{j=1}^d|b_{j1}|+\cdots+\lambda_k\sum_{j=1}^d|b_{jk}|,
\end{equation}
where $\lambda_l$ are regularization parameters whose selection will be
discussed later.
We obtain sparse principal components
by maximizing the following penalized log likelihood:
%
%
\begin{equation}\label{eq:logLp}
f(\bolds{\mu}, \mathbf{A}, \mathbf{B})  =  \ell(\bolds{\mu},
\mathbf{A}, \mathbf{B}) - n P_{\bolds{\lambda}}(\mathbf{B}),
\end{equation}
subject to the constraint that $\mathbf{A}$ has orthonormal columns.
Note that
$\mathbf{B}$ enters
the likelihood together with $\mathbf{A}$ through $\mathbf{A}\mathbf
{B}^T$ and so $\mathbf{B}$ can
be arbitrarily
small by just increasing the magnitude of $\mathbf{A}$ and not
changing the likelihood.
The orthonormal constraint on $\mathbf{A}$ prevents elements of
$\mathbf{A}$ becoming
arbitrarily large
and thus validates our use of the $L_1$ penalty on $\mathbf{B}$.

The sparse principal components can be equivalently formulated as minimizing
the following criterion function:
%
%
\begin{equation}\label{eq:logLp1}
S(\bolds{\mu}, \mathbf{A}, \mathbf{B})  =  - \ell(\bolds{\mu},
\mathbf{A}, \mathbf{B}) + n P_{\bolds{\lambda}}(\mathbf{B}),
\end{equation}
subject to the constraint that $\mathbf{A}$ has orthonormal columns.
In \eqref{eq:logLp1} the negative log likelihood can be interpreted as
a loss
function and the $L_1$ penalties increase the loss for nonzero
elements of $\mathbf{B}$ according to their magnitude. This penalized loss
interpretation is also appealing in the sense that the
independent Bernoulli trials assumption for obtaining the likelihood
\eqref{eq:logL}
need not be a realistic representation of the actual data generating process
but rather a device for generating a suitable loss function. Since the
$L_1$ penalties regularize the loss minimization, the sparse logistic
PCA is sometimes also referred to as the regularized logistic PCA. We
shall focus
on the minimization problem \eqref{eq:logLp1} for the rest of the paper.
A computational algorithm for solving the minimization problem is
presented in
Section~\ref{sec:MM}.

\begin{figure}

\includegraphics{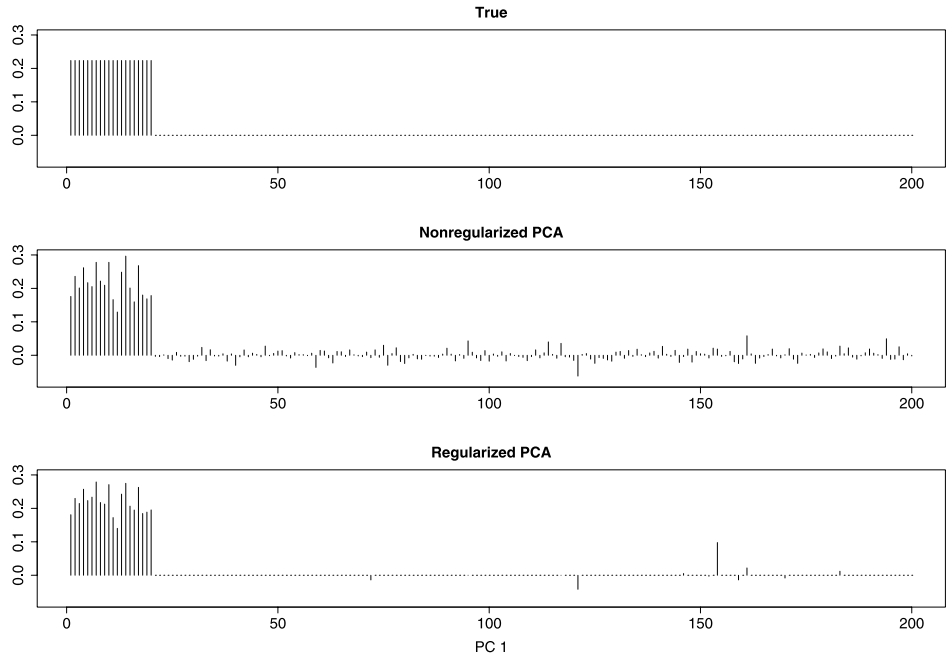}

\caption{A simulated data set with $n=100$, $d=200$ and $k=1$. Top,
middle and bottom panels show respectively the true loadings,
loadings from the nonregularized logistic PCA and from the
regularized logistic PCA. The penalty parameter is selected using the BIC.
}\label{fig:sim_loading}
\end{figure}

The effectiveness of the proposed sparse logistic PCA is illustrated in
Figure \ref{fig:sim_loading} using a rank-one model (i.e., $k=1$).
While the sparse logistic
PCA can recover the original loading vector well, the nonregularized logistic
PCA gives more noisy results. A systematic simulation study is
reported in Section \ref{sec:simu}.

\subsection{Choosing the penalty parameters}\label{sec:penalty}
Although different penalty parameters can be used for different PC loading
vectors for maximal flexibility of the methodology, we consider using
only a single penalty parameter $\lambda$ for all PC loadings. This
simplification substantially reduces the computation time, especially when
$k$ is large. Note that a larger value of $\lambda$ will lead to a smaller
number of nonzeros in the loading matrix $\mathbf{B}$ and reduced
model complexity,
but the reduced model complexity is usually associated with less good
fit of the
model. To compromise the goodness of fit and model complexity, for fixed
$k$, we choose $\lambda$ by minimizing the following BIC criterion:
%
%
\begin{equation}\label{eq:bic}
\operatorname{BIC}(\lambda) = -2\ell(\bolds{\mu},\mathbf{A},\mathbf
{B}) + \log n \times m(\lambda),
\end{equation}
where $m(\lambda)$ is a measure of the degrees of freedom.
Note that Zou, Hastie and Tibshirani (\citeyear{ZHT07}) showed that the number of nonzero coefficients
is an unbiased estimate of the degrees of freedom for the LASSO regression.
The degrees of freedom $m(\lambda)$ used in \eqref{eq:bic} is defined as
$m(\lambda)= d+nk+|\mathcal{B}(\lambda)|$, where $d$ is the length
of the vector $\bolds{\mu}$, $nk$ is the total number of elements of
$\mathbf{A}$, and
$|\mathcal{B}(\lambda)|$ is the cardinality of the index set
$\mathcal{B}(\lambda)$ of the nonzero loadings in $\mathbf{B}$ when
the penalty
parameter is $\lambda$. We use a grid search to find the optimal
$\lambda$
that minimizes the BIC.

\subsection{Determining the dimensionality of the subspace}\label{sec:k}
The BIC criterion defined in \eqref{eq:bic} can also be used to select
a suitable ``$k$.'' A two-dimensional grid search can be used to find
the minimizer of the BIC with respect to both $k$ and~$\lambda$.
To expedite computation, we implement the following strategy:
First fix $k$ at a reasonable large value and select a good $\lambda$,
then using this $\lambda$ we refine the choice of~$k$ and, finally, we refine
$\lambda$ with the refined $k$.
When optimizing with respect to~$\lambda$, a coarse grid can be used in
the first step and a finer grid in the second step. Our simulation study
showed that this strategy works reasonably well (see Section~\ref
{sec:simu-res}).\looseness=1

\begin{remark} In classical multivariate analysis, the percentage of total
variance explained by the principal components provides an intuitive
measure that can be used for subjectively choosing the appropriate number
of principal components. Zou, Hastie and Tibshirani (\citeyear{ZHT06}) and Shen and Huang
(\citeyear{SH08}) extended it to sparse PCA by modifying the definition
of variance explained by the PCs. Since there is no clear definition
of total variance for the binary data, extension of the notion of ``percentage
of variance explained'' to logistic PCA is an interesting but unsolved
problem.
\end{remark}

\section{Application to single nucleotide polymorphism data} \label{sec:snps}

Association studies based on high-throughput single nucleotide
polymorphism (SNP) data\break [Brookes~(\citeyear{Br99}); Kwok et al. (\citeyear{KDZTN96})]
have become a popular way to detect genomic regions associated
with human complex diseases. A SNP is a single base pair position
in genomic DNA at which the sequence (alleles) variation occurs
between members of a species, wherein the least frequent allele
has an abundance of 1\% or greater. A crucial issue in association
studies is population stratification detection [Hao et al. (\citeyear{HLRW04})],
which is to determine whether a population is homogeneous or has
hidden structures within it. With the presence of population stratification,
the naive case-control approach not accounting for this factor
would yield biased results [Ewens and Spielman (\citeyear{ES95})] and, therefore, draw
inaccurate scientific conclusions. See Liang and Kelemen (\citeyear{LK08})
for an extensive discussion of statistical methods and difficulties
for SNP data analysis.

The proposed sparse logistic PCA method can be used for population
stratification detection. For the purpose of demonstration, we
use the SNP data set available in the International HapMap project
[The International HapMap Consortium (\citeyear{IHP05})]. It consists of 3
different ethnic populations of 90 Caucasians (Utah residents with
ancestry from northern and western Europe; CEO),
90 Africans (Yoruba in Ibadan, Nigeria; YRI) and 90 Asians
(45 Han Chinese in Beijing, China; CHB and 45 Japanese in Tokyo, Japan; JPT).
Our task is to detect this three-subpopulation structure using the
SNP data on the 270 subjects. At many SNP locations, heterozygosity
distribution and allele frequency are known to be different among
populations and could confound the effect of the risk of disease.
To account for this factor, Serre et al. (\citeyear{SMPEYKHA08}) selected 1536 SNPs
with similar heterozygosity distribution and allele frequency.
The locations of these SNPs cover all the chromosomes except for the
sex-determining chromosome. Among these 1536~SNPs, 1392 are shared
by three ethnic groups, which are used in our analysis. We coded 0 for
the most prevalent homogeneous base pair (wild-type) and~1 for others
(mutant), resulting in a $270 \times1392$ binary matrix. This data
matrix has~2.37\% missing entries.

We applied the sparse logistic PCA to this SNP data set to explore variability
among high dimensional SNP variables, using the computation algorithm
given in Sections \ref{sec:MM} and \ref{sec:missing} below. The method
described in
Section \ref{sec:k} was used for model selection. Specifically, we initially
fixed the reduced dimension to $k=30$ and chose the penalty parameter
$\lambda$
among the rough grid of $0, 1.5^{-18}, 1.5^{-17}, \ldots, 1.5^{-10}$ using
the BIC criterion defined in Section \ref{sec:k}. Given the selected
$\lambda=1.5^{-16}$, the dimension $k$ was refined by minimizing the~BIC,
giving $k=10$. Finally, with $k=10$, we refined $\lambda$ by searching
over the grid $0, 0.0005, 0.0010, 0.0015, \ldots, 0.0100$, resulting in
$\lambda=0.0015$. As a comparison, we also applied the nonregularized logistic
PCA to the data, which corresponds to $\lambda=0$ in our general formulation
of regularized logistic PCA.

To examine which principal components represent the variability associated
with three racial groups, we used a $F$-test where scores for each fixed
PC is
regressed on the group dummy variables. For the sparse logistic PCA,
only the first two PCs were highly significant with both $p$-values
less than $0.0001$ and the remaining eight PCs were not significant with
large $p$-values (0.7681, 0.9109, 0.4764, 0.5523, 0.3376, 0.5415, 0.4480,
0.6441 for the third to the tenth PCs respectively). This result suggests
that the sparse logistic PCA can effectively compress the racial group
information into two leading PCs. Similar compression was not achieved
by the nonregularized logistic PCA; the $F$-test was significant for all the
first ten PCs with $p$-values $<$0.0001, $<$0.0001, 0.0002, 0.0001,
$<$0.0001, $<$0.0001, $<$0.0001, 0.0028, $<$0.0001 and 0.0299 respectively.

\begin{figure}

\includegraphics{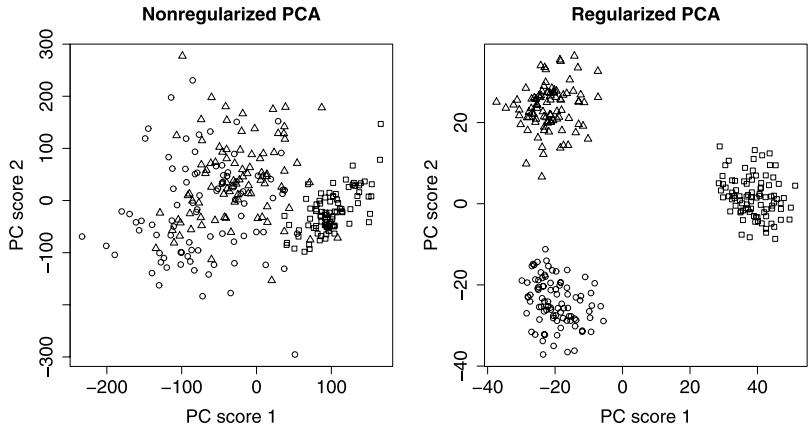}

\caption{The scatterplots of the first two PC scores from the
nonregularized (left)
and regularized logistic PCA. Circle, rectangle and triangle represent
Caucasian, African and Asian population respectively.}\label{fig:snp}
\end{figure}

Pairwise scatterplots were used to check clustering of subjects using
the PC scores.
Figure \ref{fig:snp} shows the scatterplots of first 2 PC scores with
and without regularization. The three ethnic groups are clearly separated
by the regularized PCA but not by the nonregularized PCA. To verify that
the group separation obtained is not because of luck, we permuted
observations for each SNP and applied the sparse logistic PCA
to the permuted data set; no clear clustering showed up in the PC scores.

The proposed sparse PCA method allows directly identifying the SNPs that
contribute to the group separation. The selected model has 790 and 658
nonzero loadings (representing the SNPs) respectively for the first 2 PCs,
among which 509 SNPs are shared. Therefore, 939 SNPs involved in the
first 2 PC directions are claimed to be associated with the ethnic
group effect.
Our result suggests that the population stratification factor should be
taken into
consideration at these 939~SNP locations in the subsequent study of
the association between SNPs and the disease phenotype to avoid biased
conclusion. Although in light of our simulation results, some selected SNPs
could be false positives, we believe that a large proportion of the
selected SNPs
are relevant in differentiation among the three racial groups, because
the studied SNPs were delicately selected to represent the most genetic
diversity of the whole genome [Serre et al. (\citeyear{SMPEYKHA08})]
and the genetic differentiation is the
greatest when defined on a continental basis, which is the case for our
comparison between Caucasian, Asian and African [Risch et al. (\citeyear{Risch02})].

\begin{figure}[b]

\includegraphics{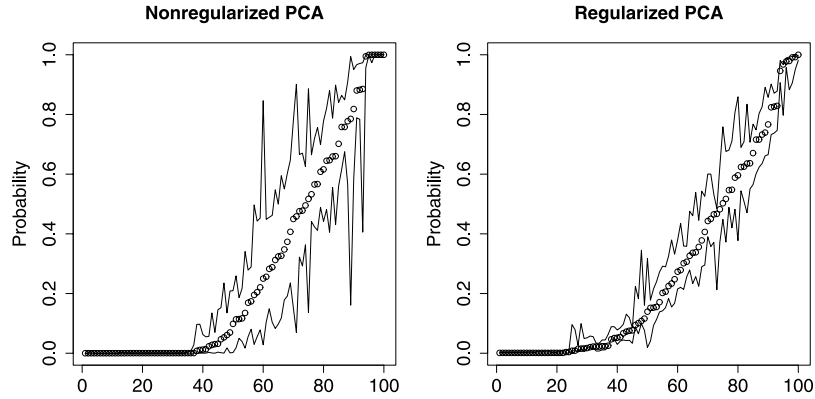}

\caption{The SNP data: 90\% bootstrap variability envelope (showed as lines)
of the probability estimates, using 100 randomly selected SNPs. Circles
are the
estimated probabilities $\hat\pi_{ij}$ from the SNP data. Results are
based on 100 bootstrap
samples.}\label{fig:boot}
\end{figure}

We further compared the regularized and nonregularized logistic PCA by assessing
the variability of the probability estimates using the parametric bootstrap.
For each method, we generated 100 bootstrapped data sets of binary matrices;
each binary matrix has entries that are independently drawn from the Bernoulli
distribution with success probability $\hat\pi_{ij}$ for the
$(i,j)$th entry,
where $\hat\pi_{ij}$ is the estimated probability. We then applied
the method
to these bootstrapped data sets to obtain 100 bootstrapped
probabilities for
each $(i,j)$ combination and to construct a 90\% variability interval using
the 5\% and 95\% quantiles of the bootstrapped probabilities. These
90\% variability
intervals were plotted against the ordered $\hat\pi_{ij}$ to form a
variability
envelop.
The variability envelop for the regularized PCA is narrower than that for
the nonregularized PCA, indicating that regularization indeed reduces
the variability of the probability estimates (Figure \ref{fig:boot}).

\begin{figure}

\includegraphics{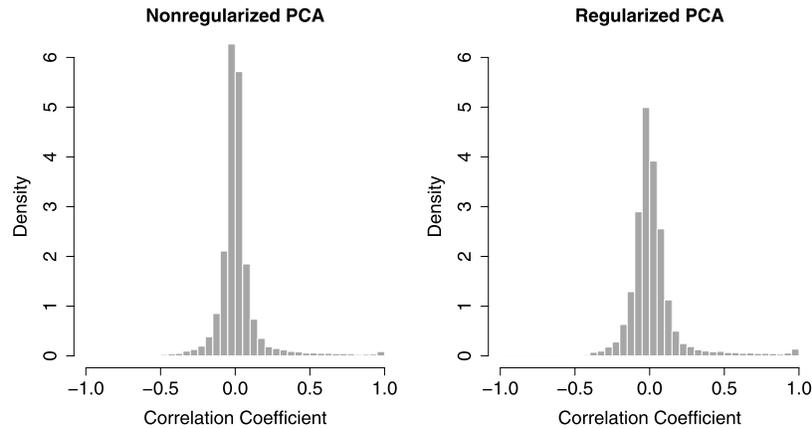}

\caption{Histograms of pairwise correlations of Pearson's residuals
from nonregularized (left) and regularized (right) logistic PCA.}\label
{fig:pairwise}
\end{figure}

Our working model for the logistic PCA specified by \eqref{eq:lik} and
\eqref{eq:model}
assumes that, conditional on the principal component scores, the observations
are independent. Since there exists spatial dependency among SNPs, one
may have concerns about the validity of our analysis results if the
dependence is strong.
In our data set, the 1536 SNPs were selected from the whole genome to capture
most of the genetic diversity in population considering factors of
physical distances, allele frequencies and linkage disequilibrium patterns.
The selected SNPs are sufficiently well separated within each chromosome
so that they can be representative of the whole genome [Serre et al. (\citeyear{SMPEYKHA08})].
Therefore, we expect that the spatial dependency in this data set
should not
be too serious to invalidate our results. To address this issue
empirically, we
first computed Pearson's residuals after fitting the models for the
nonregularized
and regularized logistic PCA, then calculated pairwise correlations of
these Pearson's
residuals for all SNP pairs for each chromosome.
Figure \ref{fig:pairwise} shows the histogram of the pairwise correlations
for each model. For both models most pairwise correlations are close to
zero, indicating that the SNPs are weakly correlated. We noticed that there
exists a very small proportion of SNP pairs that are highly correlated.
Examination
of the physical locations revealed that those highly correlated SNP
pairs consist of SNPs
in close vicinity, indicating the imperfection of the initial SNP
selection process.

\section{Computational algorithm}\label{sec:MM}

We develop a Majorization--Minimiza-\break tion (MM) algorithm for minimizing
\eqref{eq:logLp1}, which iteratively minimizes a suitably defined
quadratic upper bound of \eqref{eq:logLp1}. Instead of directly dealing
with the nonquadratic log likelihood and the nondifferentiable
sparsity inducing $L_1$ penalty, the MM algorithm sequentially
optimizes a quadratic surrogate objective function. A function
$g(x|y)$ is said to majorize a function $f(x)$ at $y$ if
\[
g(x|y)  \geq  f(x)\qquad  \mbox{for all } x\quad
 \mbox{and}\quad   g(y|y)  =  f(y).
\]
In the geometrical view the function surface $g(x|y)$ lies above
the function $f(x)$ and is tangent to it at the point $y$ so
$g(x|y)$ becomes an upper bound of $f(x)$. To minimize $f(x)$,
the MM algorithm starts from an initial guess $x^{(0)}$ of $x$,
and iteratively minimizes $g(x|x^{(m)})$ until convergence,
where $x^{(m)}$ is the estimate of $x$ at the $m$th iteration.
The MM algorithm decreases the objective function in each step
and is guaranteed to converge to a local minimum of $f(x)$. When
applying the MM algorithm, the majorizing function $g(x|y)$
is chosen such that it is easier to minimize than the original
objective function $f(x)$. See Hunter and Lange (\citeyear{HL04}) for an
introductory description of the MM algorithm.

To find a suitable majorizing function of \eqref{eq:logLp1}, we
treat the log likelihood term and the penalty term separately.
For the log likelihood term, note that, for a given point $y$,
%
\begin{eqnarray}
\quad - \log\pi(x) &\leq&- \log\pi(y) - \{1-\pi(y)\}(x-y) + \frac{2\pi
(y)-1}{4y}(x-y)^2
\label{eq:tight} \\
&\leq&-\log\pi(y) - \{1-\pi(y)\}(x-y) + \frac{1}{8}(x-y)^2, \label
{eq:uniform}
\end{eqnarray}
and the equalities hold when $x=y$ [Jaakkola and Jordan (\citeyear{JJ00});
de Leeuw (\citeyear{DeL06})]. These inequalities provide quadratic upper
bounds for the negative log inverse logit function at the
tangent point $y$. We refer to the former
bound as the tight bound, and the latter bound as the
uniform bound since its curvature does not change with $y$.
We pursue here the MM algorithm by using the uniform bound and
leave the discussion of using the tight bound to the
supplemental article [Lee, Huang and Hu (\citeyear{lee10})].
Use of the tight bound usually leads to a smaller number of iterations
of the algorithm but longer computation time because of the
complexity involved in computing the bound.
For the penalty term, the inequality
%
%
\begin{equation}\label{eq:abs}
|x|\leq\frac{x^2+y^2}{2|y|},\qquad   y\ne0,
\end{equation}
gives an upper bound for $|x|$ and the equality holds when $x=y$
[Hunter and Li (\citeyear{HL05})]. Application of \eqref{eq:uniform} and
\eqref{eq:abs} yields a suitable majorizing function of~\eqref{eq:logLp1}
and thus an MM algorithm.

Now we present details of the MM algorithm via the uniform bound.
Let $\bolds{\Theta}^{(m)}$ be the estimate of $\bolds{\Theta}$
obtained in
the $m$th step of the algorithm, with the entries $\theta_{ij}^{(m)} =
\mu_j^{(m)}+\mathbf{a}_i^{(m)T}\mathbf{b}_j^{(m)}$. By completing
the square,
the uniform bound \eqref{eq:uniform} can be rewritten as
%
%
\begin{equation}\label{eq:uniform2}
- \log\pi(x) \leq-\log\pi(y) + \tfrac{1}{8} [x-y -4 \{1-\pi(y)\}]^2.
\end{equation}
Substituting $x$ and $y$ with $q_{ij}\theta_{ij}$ and $q_{ij}\theta
_{ij}^{(m)}$
respectively in \eqref{eq:uniform2} and noticing that $q_{ij} = \pm1$,
we obtain
%
%
\begin{equation}\label{eq:uniform3}
- \log\pi(q_{ij} \theta_{ij}) \leq- \log\pi\bigl(q_{ij} \theta_{ij}^{(m)}\bigr)
+ w_{ij}^{(m)}\bigl(\theta_{ij}-x_{ij}^{(m)}\bigr)^2,
\end{equation}
where $w_{ij}^{(m)}=1/8$ and
%
%
\begin{eqnarray}\label{eq:x}
x_{ij}^{(m)} = \theta_{ij}^{(m)}+4q_{ij}\bigl\{1-\pi\bigl(q_{ij}\theta
_{ij}^{(m)}\bigr)\bigr\}.
\end{eqnarray}
The superscript $m$ of $w_{ij}^{(m)}$ and $x_{ij}^{(m)}$ indicates
the dependence on $\bolds{\Theta}^{(m)}$.\vspace*{1.5pt} Summing over all $i$, $j$ of
\eqref{eq:uniform3} and ignoring a constant term that does not depend
on unknown parameters, we obtain the following quadratic upper bound
of the negative log-likelihood:
%
%
\begin{equation}\label{eq:lik-bd}
\sum_{i=1}^n\sum_{j=1}^d w_{ij}^{(m)}\bigl(\theta_{ij}-x_{ij}^{(m)}\bigr)^2
= \sum_{i=1}^n\sum_{j=1}^d w_{ij}^{(m)}\bigl\{x_{ij}^{(m)}-(\mu
_j+\mathbf{a}
_i^T\mathbf{b}_j)\bigr\}^2.
\end{equation}
On the other hand, \eqref{eq:abs} implies that the penalty $P_{\bolds
{\lambda}
}(\mathbf{B})$
has the following quadratic upper bound:
%
%
\begin{equation}\label{eq:pen-bd}
P_{\bolds{\lambda}}(\mathbf{B}) \leq
\lambda_1 \sum_{j=1}^d \frac{b_{j1}^2 + b_{j1}^{(m)2}}{2 |b_{j1}^{(m)}|}
+ \cdots+ \lambda_k \sum_{j=1}^d \frac{b_{jk}^2 +
b_{jk}^{(m)2}}{2|b_{jk}^{(m)}|}.
\end{equation}
Combining \eqref{eq:lik-bd} and \eqref{eq:pen-bd} yields the
following quadratic upper bound (up to a constant) of the criterion function
$S(\bolds{\mu},\mathbf{A},\mathbf{B})$ defined in \eqref{eq:logLp1}:
%
%
\begin{eqnarray}\label{eq:majorizing}
&& g\bigl(\bolds{\mu},\mathbf{A},\mathbf{B}|\bolds{\mu}^{(m)},\mathbf
{A}^{(m)},\mathbf{B}^{(m)}\bigr)
\nonumber
\\[-8pt]
\\[-8pt]
\nonumber
&&\qquad  = \sum_{i=1}^{n}\sum_{j=1}^{d}
\bigl[w_{ij}^{(m)}\bigl\{x_{ij}^{(m)}-(\mu_j+\mathbf{a}_i^T\mathbf
{b}_j)\bigr\}^2
+ \mathbf{b}_j^T\mathbf{D}_{\bolds{\lambda},j}^{(m)}\mathbf
{b}_j\bigr],
\end{eqnarray}
where $\mathbf{D}_{\bolds{\lambda},j}^{(m)}$ is a diagonal matrix
with diagonal
elements $\lambda_l/\{2|b_{jl}^{(m)}|\}$ for $l=1,\ldots,k$.

\begin{theorem}\label{thm:MM}
\textup{(i)} Up to a constant that depends on $\bolds{\mu}^{(m)}$, $\mathbf
{A}^{(m)}$
and $\mathbf{B}^{(m)}$ but not on $\bolds{\mu}$, $\mathbf{A}$ and
$\mathbf{B}$, the
function $g(\bolds{\mu},\mathbf{A},\mathbf{B}|\bolds{\mu
}^{(m)},\mathbf{A}^{(m)},\mathbf{B}^{(m)})$ defined
in \eqref{eq:majorizing} majorizes $S(\bolds{\mu},\mathbf
{A},\mathbf{B})$ at
$(\bolds{\mu}^{(m)},\mathbf{A}^{(m)},\mathbf{B}^{(m)})$.

\textup{(ii)} Let $(\bolds{\mu}^{(m)},\mathbf{A}^{(m)},\mathbf{B}^{(m)})$,
$m =1, 2, \ldots,$ be
a sequence obtained by iteratively minimizing the majorizing
function. Then $S(\bolds{\mu}^{(m)},\mathbf{A}^{(m)},\mathbf
{B}^{(m)})$ decreases as $m$
gets larger and it converges to a local minimum of $S(\bolds{\mu
},\mathbf{A},\mathbf{B})$
as $m$ goes to infinity.
\end{theorem}

The majorizing function given in \eqref{eq:majorizing} is quadratic in
each of $\bolds{\mu}$, $\mathbf{A}$ and $\mathbf{B}$ when the other
two are
fixed and, thus, alternating minimization of \eqref{eq:majorizing} with
respect to $\bolds{\mu}$, $\mathbf{A}$ and $\mathbf{B}$ has closed-form
solutions, which are given below. We now drop the superscript
in $x_{ij}^{(m)}$ for notational convenience.\vspace*{-1pt} Recall that
$w_{ij}^{(m)} = 1/8$ is a constant. For fixed $\mathbf{A}$ and
$\mathbf{B}$, set
$x_{ij}^\dag=x_{ij}-\mathbf{a}_i^T\mathbf{b}_j$, the optimal $\hat
\mu_j$ is given by
%
%
\begin{equation}\label{eq:est-mu}
 \hat{\mu}_j = \mathop{\operatorname{arg\,min}}_{\mu_j}\sum
_{i=1}^{n}(x_{ij}^\dag-\mu
_j)^2
= \frac{1}{n}\sum_{i=1}^n x_{ij}^\dag,\qquad  j=1,\ldots,d.
\end{equation}
This leads to a simple matrix formula $\hat{\bolds{\mu}}=\frac
{1}{n}\mathbf{X}^{\dag
T}\mathbf{1}_n$,
which is obtained by taking the column means of $\mathbf{X}^\dag
=(x_{ij}^\dag)$.

To update $\mathbf{A}$ and $\mathbf{B}$ for fixed $\bolds{\mu}$, set
$x_{ij}^\ast=x_{ij}-\mu_j$ or in matrix form,
$\mathbf{X}^\ast= (x_{ij}^\ast) = \mathbf{X}-\mathbf{1}_n\otimes
\bolds{\mu}^T$.
Denote the $i$th row vector of $\mathbf{X}^\ast$
as $\mathbf{x}_i^{\ast T}$. For fixed $\bolds{\mu}$ and $\mathbf{B}$,
the $i$th row of $\mathbf{A}$ is updated by minimizing with respect to
$\mathbf{a}_i$ the sum of squares
$\sum_{j=1}^{d} (x_{ij}^\ast-\mathbf{a}_i^T\mathbf{b}_j)^2
= (\mathbf{x}_i^\ast-\mathbf{B}\mathbf{a}_i)^T (\mathbf{x}_i^\ast
-\mathbf{B}\mathbf{a}_i)$,
which has a closed form solution
%
%
\begin{equation}\label{eq:est-A}
\hat{\mathbf{a}}_i = (\mathbf{B}^T\mathbf{B}
)^{-1}\mathbf{B}^T\mathbf{x}_i^\ast,
 \qquad i=1,\ldots,n,
\end{equation}
or $\hat{\mathbf{A}}=\mathbf{X}^\ast\mathbf{B}(\mathbf
{B}^T\mathbf{B})^{-1}$ in matrix form.
The columns of updated $\mathbf{A}$ can be made orthonormal by
using the QR decomposition. Denote the $j$th column vector of~$\mathbf
{X}^\ast$
as~$\tilde{\mathbf{x}}_j^*$. For fixed
$\bolds{\mu}$ and $\mathbf{A}$, the $j$th row of $\mathbf{B}$ is
updated by solving
the ridge regression problem that minimizes with respect to
$\mathbf{b}_j$ the penalized sum of squares
\begin{eqnarray*}
&&\frac{1}{8}\sum_{i=1}^{n}(x_{ij}^\ast-\mathbf{a}_i^T\mathbf
{b}_j)^2 +
n\sum_{l=1}^{k}\lambda_l \frac{b_{jl}^2}{2|b_{jl}^{(m)}|}\\
&&\qquad= \frac{1}{8} (\tilde{\mathbf{x}}_j^\ast-\mathbf{A}\mathbf
{b}_j)^T (\tilde{\mathbf{x}}_j^\ast-\mathbf{A}\mathbf{b}_j)
+ n\mathbf{b}_j^T\mathbf{D}_{\bolds{\lambda},j}\mathbf{b}_j,
\end{eqnarray*}
which has a closed form solution
%
%
\begin{equation}\label{eq:est-B}
\hat{\mathbf{b}}_j = (\mathbf{A}^T\mathbf{A}+8n\mathbf
{D}_{\bolds{\lambda},j})^{-1}
\mathbf{A}^T\tilde{\mathbf{x}}_j^\ast,\qquad  j=1,\ldots,d.
\end{equation}
Since, during the iteration, $\mathbf{A}$ is made orthonormal,
$\mathbf{A}^T\mathbf{A}$
becomes the identity matrix of size $k$. Therefore, since the matrices
to be inverted are diagonal matrices, $\hat{\mathbf{b}}_j$ can be obtained
by component-wise shrinkage
\[
\hat{b}_{jl} = \frac{|b_{jl}^{(m)}|}{|b_{jl}^{(m)}|+4n\lambda_l}
\tilde
{\mathbf{a}}_l^T\tilde{\mathbf{x}}_j^*,\qquad
   l=1,\ldots,k,  j=1,\ldots,d,
\]
where $\tilde{\mathbf{a}}_l$ is the $l$th column of $\mathbf{A}$.

The MM algorithm will alternate between \eqref{eq:est-mu}, \eqref{eq:est-A}
and \eqref{eq:est-B} until convergence. The details are summarized in
Algorithm \hyperref[alg1]{1}.
In this algorithm, $k$, the number of columns of $\mathbf{A}$ and
$\mathbf{B}$,
should be specified
in advance. Different from the sequential extraction approach of Shen
and Huang (\citeyear{SH08}),
the matrices $\mathbf{A}$ and $\mathbf{B}$ obtained after applying
Algorithm \hyperref[alg1]{1} depend
on the value of
$k$, but the results are reasonably stable when $k$ is large enough.
See Section \ref{sec:k}
for discussion on choice of $k$.
We use random initial values for $\bolds{\mu}$, $\mathbf{A}$ and
$\mathbf{B}$. As with any
nonlinear optimization algorithms,
our algorithm is not guaranteed to converge to a global minimum. We can
follow the common
practice to random start the algorithm several times and find the best solution.
Our experience is that the algorithm with different initial values
usually converges to the
same solution (within the precision specified by the convergence criterion).

\begin{algorithm}[(Sparse logistic PCA algorithm I)]\label{alg1}
\begin{enumerate}
\item Initialize with $\bolds{\mu}^{(1)} = (\mu_1^{(1)}, \dots, \mu
_d^{(1)})^T$,
$\mathbf{A}^{(1)}=(\mathbf{a}_1^{(1)},\ldots,\mathbf{a}_n^{(1)})^T$ and
$\mathbf{B}^{(1)}= (\mathbf{b}_1^{(1)}, \dots, \mathbf
{b}_d^{(1)})^T $. Set $m=1$.\vspace*{1pt}
\item Compute $x_{ij}^{(m)}$ using \eqref{eq:x} and set $\mathbf
{X}^{(m)} =
(x_{ij}^{(m)})$.\vspace*{1pt}
\item Set $\mathbf{X}^{(m)\dag} = (x_{ij}^{(m)\dag})$ with
$x_{ij}^{(m)\dag} = x_{ij}^{(m)}-\mathbf{a}_i^{(m)T}\mathbf
{b}_j^{(m)}$. Update
$\bolds{\mu}$ using $\bolds{\mu}^{(m+1)}=\frac{1}{n}\mathbf
{X}^{(m)\dag T}\mathbf{1}_n$.
\item
Set $\mathbf{X}^{(m+1)\ast}= \mathbf{X}^{(m)}-\mathbf{1}_n\otimes
\bolds{\mu}^{(m+1)T}$.
\item
Update $\mathbf{A}$ by $\mathbf{A}^{(m+1)} = \mathbf
{X}^{(m+1)*}\mathbf{B}^{(m)}
(\mathbf{B}^{(m)T}\mathbf{B}^{(m)})^{-1}$.
Compute the QR decomposition $\mathbf{A}^{(m+1)} = \mathbf{Q}\mathbf
{R}$ and then replace
$\mathbf{A}^{(m+1)}$ by $\mathbf{Q}$.
\item
Set $\mathbf{C}^{(m+1)} = (c_{jl}^{(m+1)})= \mathbf
{X}^{(m+1)*T}\mathbf{A}^{(m+1)}$. Update
$\mathbf{B}$ by
$\mathbf{B}^{(m+1)} = (b_{jl}^{(m+1)})$ where
\[
{b}_{jl}^{(m+1)} = \frac{|b_{jl}^{(m)}|}{|b_{jl}^{(m)}|+4n\lambda_l}
c_{jl}^{(m+1)},
 \qquad  l=1,\ldots,k,  j=1,\ldots,d.
\]
\item Repeat steps 2 through 6 with $m$ replaced by $m+1$ until convergence.
\end{enumerate}\vspace*{-8pt}
\end{algorithm}

\begin{remark} The orthogonalization in step 5 of Algorithm \hyperref[alg1]{1} does not change
the descent property of the MM algorithm. Let $A^{(m+1)}$ be the optimizer
before orthogonalization. Then $S(A^{(m+1)}, B^{(m)}) \leq S(A^{(m)}, B^{(m)})$,
where, for simplicity, $\mu$~is omitted from the objective function $S$.
Let $A^{(m+1)} = \widetilde{A}^{(m+1)} R$ be the QR decomposition of
$A^{(m+1)}$
and let $\widetilde{B}^{(m)} = B^{(m)} R^T$. Then
$\widetilde{A}^{(m+1)} \widetilde{B}^{(m)T} = A^{(m+1)} B^{(m)T}$ and so
$S(\widetilde{A}^{(m+1)}, \widetilde{B}^{(m)}) = S(A^{(m+1)}, B^{(m)})$.
Consequently, $S(\widetilde{A}^{(m+1)}, \widetilde{B}^{(m)}) \leq
S(A^{(m)}, B^{(m)})$.
\end{remark}

\section{Handling missing data}\label{sec:missing}

Missing data are commonly encountered in real applications.
In this section we extend our sparse logistic PCA method to cases when
missing data are present.

Let $\mathcal{N}=\{(i,j)|y_{ij} \mbox{ is not observed}\}$ denote
the index set for missing values. The sparse logistic PCA
minimizes the following criterion function:
%
%
\begin{equation}\label{eq:logLp1-miss}
T(\bolds{\mu}, \mathbf{A}, \mathbf{B})  =  - \ell_{\mathit{obs}}(\bolds
{\mu}, \mathbf{A}, \mathbf{B}) + n P_{\bolds{\lambda}}(\mathbf{B}),
\end{equation}
where
%
%
\begin{equation}
\ell_{\mathit{obs}}(\bolds{\mu}, \mathbf{A}, \mathbf{B}) = \mathop{\sum
\sum}_{(i,j)\notin\mathcal{N}}\log\pi\{
q_{ij} (\mu_j+\mathbf{a}_i^T\mathbf{b}_j)\}
\end{equation}
can be interpreted as the observed data log likelihood for model
\eqref{eq:model}. Similar to the nonmissing data case, direct
minimization of \eqref{eq:logLp1-miss} is not straightforward
because the log likelihood term is not quadratic and the penalty
term is nondifferentiable. Direct minimization of \eqref{eq:logLp1-miss}
is also complicated by the fact that the summation in the definition
of the observed data log likelihood is not over a rectangular region.
Again, we develop an iterative MM algorithm to solve the optimization
problem. The strategy is to fill in the missing data with the fitted
values based on the current parameter estimates, then proceed with the
algorithm that assumes complete data, and iterate until convergence.

Define the working variables
%
%
\begin{equation}
z_{ij}^{(m)}  =
\cases{
x_{ij}^{(m)}, &\quad  $(i,j)\notin\mathcal{N},$\vspace*{2pt} \cr
\theta_{ij}^{(m)} = \mu_j^{(m)}+\mathbf{a}_i^{(m)T}\mathbf
{b}_j^{(m)}, &\quad  $(i,j)\in
\mathcal{N},$}
\end{equation}
where $x_{ij}^{(m)}$ is defined in (\ref{eq:x}). Let
%
%
\begin{eqnarray}\label{eq:majorizing-miss}
&& h\bigl(\bolds{\mu},\mathbf{A},\mathbf{B}|\bolds{\mu}^{(m)},\mathbf
{A}^{(m)},\mathbf{B}^{(m)}\bigr)
\nonumber
\\[-8pt]
\\[-8pt]
\nonumber
&&\qquad = \sum_{i=1}^{n}\sum_{j=1}^{d}\bigl[w_{ij}^{(m)}
\bigl\{z_{ij}^{(m)}-(\mu_j+\mathbf{a}_i^T\mathbf{b}_j)\bigr\}^2
+\mathbf{b}_j^T\mathbf{D}_{\bolds{\lambda},j}^{(m)}\mathbf
{b}_j\bigr],
\end{eqnarray}
where $\mathbf{D}_{\bolds{\lambda},j}^{(m)}$ are diagonal matrices
with diagonal
elements $\lambda_l/\{2|b_{jl}^{(m)}|\}$\vspace*{1pt} for $l=1,\ldots,k$. The
following result extends Theorem \ref{thm:MM} to the missing data case.
The proof is given in the \hyperref[app]{Appendix}.

\begin{theorem}\label{thm:MM-missing}
\textup{(i)} Up to a constant that depends on $\bolds{\mu}^{(m)}$, $\mathbf
{A}^{(m)}$
and $\mathbf{B}^{(m)}$ but not on $\bolds{\mu}$, $\mathbf{A}$ and
$\mathbf{B}$, the
function $h(\bolds{\mu},\mathbf{A},\mathbf{B}|\bolds{\mu
}^{(m)},\mathbf{A}^{(m)},\mathbf{B}^{(m)})$ defined
in \eqref{eq:majorizing-miss} majorizes $T(\bolds{\mu},\mathbf
{A},\mathbf{B})$ at
$(\bolds{\mu}^{(m)},\mathbf{A}^{(m)},\mathbf{B}^{(m)})$.

\textup{(ii)} Let $(\bolds{\mu}^{(m)},\mathbf{A}^{(m)},\mathbf{B}^{(m)})$,
$m =1, 2, \dots,$
be a sequence obtained by iteratively minimizing the majorizing
function. Then $T(\bolds{\mu}^{(m)},\mathbf{A}^{(m)},\mathbf
{B}^{(m)})$ decreases as $m$
gets larger and it converges to a local minimum of $T(\bolds{\mu
},\mathbf{A},\mathbf{B})$
as $m$ goes to infinity.
\end{theorem}

Note that the majorizing functions given in \eqref{eq:majorizing-miss}
have the same form as those given in \eqref{eq:majorizing}
except that $x_{ij}^{(m)}$ in \eqref{eq:majorizing} is changed to
$z_{ij}^{(m)}$ in \eqref{eq:majorizing-miss}. Thus, the
computation algorithm developed in Section \ref{sec:MM} is readily
applicable in the missing data case with a simple replacement
of $x_{ij}^{(m)}$ by $z_{ij}^{(m)}$.\vspace*{1pt} The working variable $z_{ij}^{(m)}$
in \eqref{eq:majorizing-miss} is easily understood: It
is the same as the nonmissing data case if $y_{ij}$ is observable;
otherwise, it is an imputed $\theta_{ij}$ value based on the
reduced rank model \eqref{eq:model} and the current guess of $\bolds
{\mu}$,
$\mathbf{A}$ and $\mathbf{B}$.

\section{Simulation study}\label{sec:simu}
In this section we demonstrate our sparse logistic PCA method
using a simulation study. The method worked well in various settings
that we tested, but here we only report results in a challenging case
that the number of variables $d$ is bigger than the sample size $n$.

\subsection{The signal-to-noise ratio}
To facilitate setting up simulation studies, we introduce a notion of
signal-to-noise ratio for logistic PCA.
In our logistic PCA model, the entries of the $n\times d$ data
matrix are independent Bernoulli random variables with success
probability $\pi_{ij}=\{1+\exp(-\theta_{ij})\}^{-1}$ for the
$(i,j)$th cell. The matrix of canonical parameters $\bolds{\Theta}=
(\theta_{ij})$
has a reduced rank representation $\bolds{\Theta}= \mathbf{1}\otimes
\bolds{\mu}^T +
\mathbf{A}\mathbf{B}^T$, where $\mathbf{A}$ is a $n\times k$ matrix
of PC scores and $\mathbf{B}$ is
a sparse $d\times k$ PC loading matrix. In our simulation study, elements
of the $l$th column of $\mathbf{A}$ are independent draws from a zero-mean
Gaussian distribution with variance $\sigma_{al}^2$, $1\leq l \leq k$.
The variance $\sigma_{al}^2$ measures the signal level of the $l$th PC.
We set up the PC variances relative to a suitably defined baseline noise
level.

We define a baseline noise level for fixed $n$, $d$ and $k$ as follows.
First we create a binary data matrix by generating $n\times d$ independent
binary variables from Bernoulli distribution with the success probability
$1/2$. These binary variables are understood to come from the pure noise
since they are generated without having any structure on the success
probabilities. Then, we conduct a $k$-component logistic PCA without
regularization and compute the average of the sample variances of the
obtained $k$ PC scores, which is denoted as $\sigma_b^2$.
We repeat the above process of generating ``pure noise'' binary data
matrices a large number of times (e.g., 100) and take the mean
of $\sigma_b^2$ computed from these matrices as the baseline noise level.

With the notion of baseline noise level, we define the signal-to-noise
ratio (SNR) for a PC as
%
%
\begin{equation}
\operatorname{SNR}  =  \frac{\mbox{variance of PC scores}}{\mbox{baseline noise level}}.
\end{equation}
In our simulation study we first compute the baseline noise level for
a given combination of $n$, $d$ and $k$, then use the above formula
to specify the variances of PC scores based on the fixed values of SNR.

\subsection{Simulation setup}
We set the intrinsic dimension to be $k=2$ and the number of rows
of the data matrix to be $n=100$. We varied the number of variables~$d$ and the signal-to-noise ratio SNR.
We considered three choices of $d$: $d=200$, $d=500$ and $d=1000$.
The scores of the $l$th PC were randomly drawn from the
$N(0, \sigma^2_{al})$ distribution with
$\sigma_{al}^2 = \operatorname{SNR}_l \,\cdot\,(\mbox{baseline noise level})$,
where $\operatorname{SNR}_l$ is the SNR for the $l$th PC.
We considered two settings of SNR: $(3,2)$ and $(5,3)$.
For example, when the SNR is $(3,2)$, the variance of the first PC
is 3 times the baseline noise level and the variance of the second
PC is 2 times the baseline noise level.
We construct two sparse PC loading vectors as follows: Let $b_{j1}$
and $b_{j2}$ denote correspondingly the components of the first and
the second PC loading vectors. We let $b_{j1}=1$ for $j=1,\ldots,20$,
$b_{j2}=1$ for $j=21,\ldots,40$ and the rest of $b_{jl}$ are all
taken to be $0$. The mean vector $\bolds{\mu}$ was set to be a vector
of zeros.

\subsection{Simulation results}\label{sec:simu-res}

Logistic PCA with and without sparsity inducing regularization was
conducted on 100 simulated data sets for each setting. When applying
the sparse logistic PCA algorithm, three choice of $k$ were considered:
$k$ is fixed at the true value ($k=2$), at a moderately large value
($k=30$), and selected using the BIC. The penalty parameter was selected
using the method described in Section \ref{sec:penalty}.

To measure the closeness of the estimated PC loading matrix $\widehat
{\mathbf{B}}$
and the true loading matrix $\mathbf{B}$, we use the principal angle
between spaces spanned by $\widehat{\mathbf{B}}$ and~$\mathbf{B}$.
The principal
angle measures the maximum angle between any two vectors on
the spaces generated by the columns of $\widehat{\mathbf{B}}$ and
$\mathbf{B}$.
More precisely, it is defined by $\cos^{-1}(\rho)\times180/\pi$,
where $\rho$ is the minimum eigenvalue of the matrix
$\mathbf{Q}_{\widehat{\mathbf{B}}}^T\mathbf{Q}_{\mathbf{B}}$,
where $\mathbf{Q}_{\widehat{\mathbf{B}}}$ and
$\mathbf{Q}_{\mathbf{B}}$ are orthogonal basis matrices obtained by
the QR
decomposition of matrices $\widehat{\mathbf{B}}$ and $\mathbf{B}$,
respectively
[Golub and van Loan (\citeyear{GVL96})].

%
\begin{table}
\caption{The results of logistic PCA with
and without sparsity inducing regularization, based on 100 simulated
data sets for each setting. The reported values are the mean
(standard error) of the principal angle $(^\circ)$ between
the estimated and the true PC loading matrices}\label{tab:angle}
\begin{tabular*}{\textwidth}{@{\extracolsep{\fill}}lcccc@{}}
\hline
$\bolds{d}$ & \multicolumn{1}{c}{\textbf{SNR}} & \multicolumn{1}{c}{$\bolds{k=2}$} &
\multicolumn{1}{c}{$\bolds{k=30}$} & \multicolumn{1}{c}{\textbf{Selected} $\bolds{k}$} \\
\hline
\phantom{0}$200$ & \multicolumn{1}{l}{$\mathrm{SNR}=(3,2)$} & & & \\
& Nonregularized & $12.532 $ $(0.115)$ & $35.725 $ $(0.177)$ & --
\\
& Regularized  & \phantom{0}$5.860 $ $(0.123)$ & $10.125 $ $(0.324)$ & $5.816
$ $(0.125)$ \\[3pt]
& \multicolumn{1}{l}{$\mathrm{SNR}=(5,3)$} & & & \\
& Nonregularized & $11.913 $ $(0.122)$ & $36.350$ $(0.189)$ & -- \\
& Regularized  & \phantom{0}$5.803 $ $(0.128)$ & \phantom{0}$9.843 $ $(0.321)$ & $5.769 $
$(0.127)$ \\[6pt]
\phantom{0}$500$ & \multicolumn{1}{l}{$\mathrm{SNR}=(3,2)$} & & & \\
& Nonregularized & $10.890$ $(0.095)$ & $31.884$ $(0.188)$ & -- \\
& Regularized  & \phantom{0}$4.731 $ $(0.115)$ & \phantom{0}$9.413 $ $(0.282)$ & $4.690 $
$(0.101)$ \\[3pt]
& \multicolumn{1}{l}{$\mathrm{SNR}=(5,3)$} & & & \\
& Nonregularized & $10.166 $ $(0.095)$ & $31.941 $ $(0.193)$ & --
\\
& Regularized  & \phantom{0}$4.729 $ $(0.121)$ & \phantom{0}$9.242 $ $(0.252)$ & $4.544 $
$(0.119)$ \\[6pt]
$1000$ & \multicolumn{1}{l}{$\mathrm{SNR}=(3,2)$} & & & \\
& Nonregularized & $12.018 $ $(0.167)$ & $36.040$ $(0.181)$ & -- \\
& Regularized  & \phantom{0}$7.015 $ $(0.486)$ & $11.807 $ $(0.433)$ & $4.534
$ $(0.141)$ \\[3pt]
& \multicolumn{1}{l}{$\mathrm{SNR}=(5,3)$} & & & \\
& Nonregularized & $11.370$ $(0.156)$ & $36.144 $ $(0.180)$ & -- \\
& Regularized  & \phantom{0}$6.767 $ $(0.474)$ & $10.825 $ $(0.475)$ & $4.196$
$(0.127)$ \\
\hline
\end{tabular*}
\end{table}

The mean and standard deviation of principal angles for logistic PCA
with and without regularization are presented in Table \ref{tab:angle}.
Since smaller principal angles indicate better estimates of the PC loading
matrix, the sparsity inducing regularization has a clear benefit---it
can substantially reduce the principal angles. The benefit is even more
profound when the number of PCs used in the program ($k=30$) is larger
than the true number that was used to generate the data ($k=2$).
The performance of sparse logistic PCA with selected $k$ is similar
to that when $k$ is fixed at the true value.
Frequencies of the selected $k$ from $100$ simulation data sets in each
settings of Table \ref{tab:angle} are shown in Table \ref{tab:k}.
When $d=200$, the BIC finds well the true $k=2$ but, as $d$ gets
larger, there is a trend that a slightly larger $k$ is selected. The
performance of using BIC to select $k$ is considered as
quite good, given that the sample size is only 100.

%
\begin{table}
\caption{Frequencies of the selected $k$
using the BIC}\label{tab:k}
\begin{tabular*}{250pt}{@{\extracolsep{\fill}}lcccccccc@{}}
\hline
& & \multicolumn{7}{c@{}}{\textbf{Selected} $\bolds{k}$} \\[-6pt]
& & \multicolumn{7}{c@{}}{\hrulefill} \\
$\bolds{d}$ & \textbf{SNR} & $\mathbf{1}$ & $\mathbf{2}$ & $\mathbf{3}$ & $\mathbf{4}$ & $\mathbf{5}$ & $\mathbf{6}$ & $\mathbf{7}$ \\
\hline
\phantom{0}$200$ & $(3,2)$ & $0$ & $95$ & \phantom{0}$5$ & \phantom{0}$0$ & \phantom{0}$0$ & $0$ & $0$ \\
& $(5,3)$ & $0$ & $96$ & \phantom{0}$4$ & \phantom{0}$0$ & \phantom{0}$0$ & $0$ & $0$ \\[3pt]
\phantom{0}$500$ & $(3,2)$ & $1$ & $58$ & $37$ & \phantom{0}$4$ & \phantom{0}$0$ & $0$ & $0$ \\
& $(5,3)$ & $0$ & $60$ & $36$ & \phantom{0}$3$ & \phantom{0}$1$ & $0$ & $0$ \\[3pt]
$1000$ & $(3,2)$ & $3$ & $34$ & $36$ & $15$ & $10$ & $1$ & $1$ \\
& $(5,3)$ & $2$ & $31$ & $47$ & $15$ & \phantom{0}$4$ & $1$ & $0$ \\
\hline
\end{tabular*}
\end{table}
%

%
\begin{table}[b]
\caption{The results of logistic PCA with sparsity
inducing regularization, based on 100 simulated data sets for each setting
in Table \textup{\protect\ref{tab:angle}}. The reported values are the mean (standard
error) of
the percentages of false positives. The description of results is in
the text}
\label{tab:f-postive}
\begin{tabular*}{\textwidth}{@{\extracolsep{\fill}}lcccc@{}}
\hline
$\bolds{d}$ & \textbf{SNR} & $\bolds{k=2}$ & $\bolds{k=30}$ & \textbf{Selected} $\bolds{k}$ \\
\hline
\phantom{0}$200$ & $(3,2)$ & $45.05 (1.54)$ & $41.51 \mbox{ }(1.39)$ & $44.94 \mbox{ }(1.51)$ \\
& $(5,3)$ & $48.16 \mbox{ }(1.63)$ & $40.53 \mbox{ }(1.36)$ & $48.26 \mbox{ }(1.63)$ \\
[3pt]
\phantom{0}$500$ & $(3,2)$ & $14.83 \mbox{ }(0.74)$ & $18.91 \mbox{ }(0.51)$ & $16.70 \mbox{ }(0.72)$ \\
& $(5,3)$ & $16.06 \mbox{ }(0.68)$ & $18.78 \mbox{ }(0.42)$ & $16.93 \mbox{ }(0.68)$ \\
[3pt]
$1000$ & $(3,2)$ & $10.87 \mbox{ }(0.75)$ & $12.80\mbox{ }(0.73)$ & $10.13 \mbox{ }(0.60)$ \\
& $(5,3)$ & $10.89 \mbox{ }(0.70)$ & $12.86 \mbox{ }(0.73)$ & \phantom{0}$9.26 \mbox{ }(0.50)$ \\
\hline
\end{tabular*}
\end{table}
A useful feature of the sparse logistic PCA is its ability to select
relevant variables when estimating the PC loading vectors. A zero loading
of a variable on a PC means that the corresponding variable is not used
when forming that PC, and a nonzero loading indicates a useful variable.
Our experience with simulated data shows that nonzero loadings can almost
always be identified by the method, but some identified nonzero loadings
may correspond to irrelevant variables, cases of false positives.
Table \ref{tab:f-postive} presents the percentages of false positives for
various settings reported in Table \ref{tab:angle}. When $d$ is 500 or
1000, the percentages of false positives are low, all below 20\%. But
when $d$ is 200, the percentages of false positives are between 40\% and
50\%, suggesting big room for improvement in variable selection.

\section{Discussion and extension}\label{sec7}
In this paper we propose a sparse PCA method for analyzing binary data
by maximizing a penalized Bernoulli likelihood. The sparsity inducing
$L_1$ penalty is used to acquire simple principal components for the
sake of easy interpretation and stable estimation. The MM algorithm
developed for implementation of our method provides a unified solution
for dealing with (i)~the nonquadratic likelihood, (ii)~the nondifferentiable
penalty function, and (iii) presence of missing data. Although the
theoretical derivation is not straightforward, the steps of the algorithm
are very simple---they are (weighted) penalized least squares with
closed-form expressions.

We have focused on the logit link so far, but other link functions can
also be used.
In particular, a slight modification of the proposed method can handle the
probit link, where the success probabilities $\theta_{ij} =
\Phi^{-1}(\pi_{ij})$ with $\Phi(\cdot)$ being the c.d.f. of the standard
Gaussian distribution. The log likelihood function \eqref{eq:logL} of the
reduced rank model is changed to
%
%
\begin{equation}\label{eq:logL-probit}
\ell(\bolds{\mu}, \mathbf{A}, \mathbf{B})  =  \sum_{j=1}^d \sum
_{i=1}^n\log\Phi\{q_{ij} (\mu
_j+\mathbf{a}_i^T\mathbf{b}_j)\}.
\end{equation}
Instead of using the majorization in \eqref{eq:uniform}, we apply the following
upper bound to majorize the negative log likelihood:
%
%
\begin{equation}\label{eq:major-probit}
- \log\Phi(x) \leq- \log\Phi(y) - \frac{\phi(y)}{\Phi(y)}(x-y) +
\frac{1}{2} (x-y)^2,
\end{equation}
where $\phi(\cdot)$ is the Gaussian density [B\"ohning (\citeyear{boh99}); de
Leeuw (\citeyear{DeL06})].
Algorithm~\hyperref[alg1]{1} still applies with appropriate changes to the definitions
of the
weights $w_{ij}^{(m)}$ and the working variables $x_{ij}^{(m)}$.

Our method can also be extended in a straightforward way to handle composite
data which consists of both binary and continuous variables. While the
binary variables are modeled with Bernoulli distributions, the continuous
variables can be modeled with Gaussian distributions. Including some
continuous variables corresponds to adding some negative Gaussian log
likelihood terms to the log likelihood expression \eqref{eq:logL}.
Since the Gaussian log likelihood is quadratic, it blends in easily with
the quadratic majorization used for the logistic PCA.\vspace*{1pt} Specifically,
if the $j$th variable is of a continuous type, we assume
$y_{ij} \sim N(\theta_{ij}, \sigma^2)$ with $\theta_{ij}$ satisfying~\eqref{eq:model},
and simply let $x_{ij}^{(m)}=y_{ij}$ and
$w_{ij}^{(m)}=1/\sigma^2$ when forming the majorizing
function \eqref{eq:majorizing}. The residual variance $\sigma^2$ of
fitting the continuous variables can be estimated using the
residual sum of squares. Taking into account the fact that
different weighting schemes are used for the binary variables and
the continuous variables in the majorizing function, a slight
modification of Algorithm 2 presented in the supplemental article
[Lee, Huang and Hu (\citeyear{lee10})] can be used for computation.

\begin{appendix}

\section*{Appendix}\label{app}

\subsection{\texorpdfstring{Proof of Theorem \protect\ref{thm:MM}}{Proof of Theorem 4.1}}\label{sec:proof-MM}
We prove the results for both the tight and the uniform bound case.
Applications of \eqref{eq:tight} and \eqref{eq:uniform} yield the
following majorizing functions of the negative log likelihood
$-\ell(\bolds{\mu},\mathbf{A},\mathbf{B})$:
\begin{eqnarray*}
&&\sum_{i=1}^n\sum_{j=1}^d \biggl[ - \log\pi\bigl(q_{ij}\theta_{ij}^{(m)}\bigr)
- q_{ij}\bigl\{1-\pi\bigl(q_{ij}\theta_{ij}^{(m)}\bigr)\bigr\}\bigl(\theta-\theta
_{ij}^{(m)}\bigr)\\
&&\qquad\hspace*{96pt}{} + \frac{2\pi(q_{ij}\theta_{ij}^{(m)}) -1}{4q_{ij}\theta
_{ij}^{(m)}}\bigl(\theta-\theta_{ij}^{(m)}\bigr)^2 \biggr]
\end{eqnarray*}
for the tight bound, and
\[
\sum_{i=1}^n\sum_{j=1}^d \biggl[ - \log\pi\bigl(q_{ij}\theta_{ij}^{(m)}\bigr)
- q_{ij}\bigl\{1-\pi\bigl(q_{ij}\theta_{ij}^{(m)}\bigr)\bigr\}\bigl(\theta-\theta_{ij}^{(m)}\bigr)
+ \frac{1}{8}\bigl(\theta-\theta_{ij}^{(m)}\bigr)^2 \biggr]
\]
for the uniform bound. Note that
\[
\bigl\{2\pi\bigl(q_{ij}\theta_{ij}^{(m)}\bigr) -1\bigr\}/\bigl\{4q_{ij}\theta_{ij}^{(m)}\bigr\} =
\bigl\{
2\pi\bigl(\theta_{ij}^{(m)}\bigr)
-1\bigr\}/\bigl\{4\theta_{ij}^{(m)}\bigr\}
\]
for $q_{ij} = \pm1$. By completing the squares and using the
definitions of $x^{(m)}_{ij}$ and~$w_{ij}^{(m)}$, these majorizing functions can be rewritten as
\begin{eqnarray*}
&& - \tilde{\ell}\bigl(\bolds{\mu},\mathbf{A},\mathbf{B}|\bolds{\mu
}^{(m)},\mathbf{A}^{(m)},\mathbf{B}^{(m)}\bigr)\\
&&\qquad  = - \ell\bigl(\bolds{\Theta}^{(m)}\bigr) - 2 \sum_{i=1}^n\sum
_{j=1}^d \bigl\{1-\pi
\bigl(q_{ij}\theta_{ij}^{(m)}\bigr)\bigr\}^2
+ \sum_{i=1}^n\sum_{j=1}^d w_{ij}^{(m)} \bigl(\theta_{ij}-x^{(m)}_{ij}\bigr)^2.
\end{eqnarray*}
On the other hand, application of \eqref{eq:abs} yields the following
majorizing function of $P_{\bolds{\lambda}}(\mathbf{B})$:
\begin{eqnarray*}
\tilde{P}_{\bolds{\lambda}}\bigl(\mathbf{B}|\mathbf{B}^{(m)}\bigr) & =&
\lambda_1 \sum_{j=1}^{d}\frac{b_{j1}^2+b_{j1}^{(m)2}}{2|b_{j1}^{(m)}|}
+ \cdots+ \lambda_k
\sum_{j=1}^{d}\frac{b_{jk}^2+b_{jk}^{(m)2}}{2|b_{jk}^{(m)}|} \\
&=& \sum_{j=1}^{d} \mathbf{b}_j^{(m)T}\mathbf{D}_{\bolds{\lambda
},j}^{(m)}\mathbf{b}_j^{(m)} +
\sum_{j=1}^{d} \mathbf{b}_j^T \mathbf{D}_{\bolds{\lambda
},j}^{(m)}\mathbf{b}_j.
\end{eqnarray*}
Since the majorization relation between functions is closed under the
formation of sums, $- \tilde\ell+ n
\tilde{P}_{\bolds{\lambda}}(\mathbf{B}|\mathbf{B}^{(m)})$
majorizes $S(\bolds{\mu},\mathbf{A},\mathbf{B})$ at
$(\bolds{\mu}^{(m)},\mathbf{A}^{(m)},\mathbf{B}^{(m)})$. Noticing
that $-
\tilde\ell+ n \tilde{P}_{\bolds{\lambda}}(\mathbf{B}|\mathbf
{B}^{(m)})$ equals $g(\bolds{\mu},\mathbf{A}
,\mathbf{B}|\bolds{\mu}^{(m)},\mathbf{A}^{(m)},\mathbf{B}^{(m)})$
up to a constant
independent of $(\bolds{\mu},\mathbf{A},\mathbf{B})$, we complete
the proof of part (i).
Part (ii) of the theorem follows from the general property of
the MM algorithm [Hunter and Lange~(\citeyear{HL04})].
\qed

\subsection{\texorpdfstring{Proof of Theorem
\protect\ref{thm:MM-missing}}{Proof of Theorem 5.1}}
Note that the objective function to be minimized is the
summation of two terms---the log likelihood term and the penalty term.
Because the majorization property is closed under function summation,
we deal with the two terms separately. We can find a
majorization function of the penalty term as in Theorem \ref{thm:MM}.
To find a majorization function of the log likelihood term,
we apply the argument in the standard EM algorithm for handling
missing data [Dempster, Laird and Rubin (\citeyear{DLR77})]. The complete data log likelihood is
\[
\ell_{\mathit{com}}(\bolds{\mu},\mathbf{A},\mathbf{B}) = \mathop{\sum\sum
}_{(i,j)\notin\mathcal{N}}\log\pi
(q_{ij}\theta_{ij}) + \mathop{\sum\sum}_{(i,j)\in\mathcal{N}}
\log\pi(q_{ij}\theta_{ij}).
\]
Its conditional expectation given the observed data and the current
guess of the parameter values is
%
%
\setcounter{equation}{0}
\begin{eqnarray}\label{eq:EM-Q}
&& Q\bigl(\bolds{\mu},\mathbf{A},\mathbf{B}|\bolds{\mu}^{(m)},\mathbf
{A}^{(m)},\mathbf{B}^{(m)}\bigr) \nonumber\\
&&\qquad  = \mathop{\sum\sum}_{(i,j)\notin\mathcal{N}}\log\pi
(q_{ij}\theta_{ij})\\
&&\qquad\quad{}   + \mathop{\sum\sum}_{(i,j)\in\mathcal{N}}E\bigl[\log\pi
(q_{ij}\theta_{ij})
|\mathbf{Y}_o,\bolds{\mu}^{(m)},\mathbf{A}^{(m)},\mathbf{B}^{(m)}\bigr],\nonumber
\end{eqnarray}
where $\mathbf{Y}_o$ denotes the observed data. By the standard EM theory,
%
%
\begin{eqnarray}\label{eq:EM-major}
\hspace*{11pt}- \tilde\ell_{\mathit{obs}}(\bolds{\mu},\mathbf{A},\mathbf{B}) &\triangleq&- Q\bigl(\bolds{\mu},\mathbf{A},\mathbf{B}|\bolds
{\mu}^{(m)},\mathbf{A}^{(m)},\mathbf{B}^{(m)}\bigr)
- \ell_{\mathit{obs}}\bigl(\bolds{\mu}^{(m)},\mathbf{A}^{(m)},\mathbf{B}^{(m)}\bigr)
\nonumber
\\[-8pt]
\\[-8pt]
\nonumber
&&{}   + Q\bigl(\bolds{\mu}^{(m)},\mathbf{A}^{(m)},\mathbf
{B}^{(m)}|\bolds{\mu}^{(m)},\mathbf{A}
^{(m)},\mathbf{B}^{(m)}\bigr)\nonumber
\end{eqnarray}
majorizes $- \ell_{\mathit{obs}}(\bolds{\mu},\mathbf{A},\mathbf{B})$ at
$(\bolds{\mu}^{(m)},\mathbf{A}^{(m)},\mathbf{B}^{(m)})$,
that is, $ - \tilde\ell_{\mathit{obs}}(\bolds{\mu},\mathbf{A},\mathbf{B})
\geq\break -\ell_{\mathit{obs}}(\bolds{\mu},\mathbf{A}
,\mathbf{B}), $
and the equality holds when $(\bolds{\mu},\mathbf{A},\mathbf{B}) =
(\bolds{\mu}^{(m)},\mathbf{A}^{(m)},\mathbf{B}^{(m)})$.

Now we find a quadratic majorizing function of $ - \tilde\ell
_{\mathit{obs}}(\bolds{\mu}
,\mathbf{A},\mathbf{B})$,
which in turn majorizes $- \ell_{\mathit{obs}}(\bolds{\mu},\mathbf
{A},\mathbf{B})$ because of the transitivity
of the majorization relation. We need only to find a quadratic majorization
function of $-Q(\bolds{\mu},\mathbf{A},\mathbf{B}|\bolds{\mu
}^{(m)},\mathbf{A}^{(m)},\mathbf{B}^{(m)})$, since it is
the only term in the definition \eqref{eq:EM-major} of
$-\tilde\ell_{\mathit{obs}}(\bolds{\mu},\mathbf{A},\mathbf{B})$ that
depends on the unknown parameters.
According to \eqref{eq:EM-Q}, $-Q(\bolds{\mu},\mathbf{A},\mathbf
{B}|\bolds{\mu}^{(m)},\mathbf{A}^{(m)},\mathbf{B}^{(m)})$
can be decomposed into two terms, one corresponding to observed data,
the other
corresponding to the missing data. The former term can been treated as
in the
proof of Theorem \ref{thm:MM}. When
$(i,j)\notin\mathcal{N}$, $-\log\pi(q_{ij}\theta_{ij})$ is
majorized by
$w_{ij}^{(m)}(\theta_{ij} - x_{ij}^{(m)})^2$, up to a constant. To
treat the
latter term, note that, when $(i,j)\in\mathcal{N}$,
\begin{eqnarray*}
&& E\bigl[\log\pi(q_{ij}\theta_{ij})|\mathbf{Y}_o,\bolds{\mu
}^{(m)},\mathbf{A}^{(m)},\mathbf{B}
^{(m)}\bigr]\\
&&\qquad  = \pi\bigl(\theta_{ij}^{(m)}\bigr)\log\pi(\theta_{ij}) + \bigl\{1-\pi
\bigl(\theta
_{ij}^{(m)}\bigr)\bigr\}\log\{1-\pi(\theta_{ij})\}\\
&&\qquad  = \sum_{q_{ij} = \pm1} \pi\bigl(q_{ij}\theta_{ij}^{(m)}\bigr) \log
\pi
(q_{ij}\theta_{ij}),
\end{eqnarray*}
using the fact that the missing data are independent of the observed data,
and that $1-\pi(\theta) = \pi(-\theta)$. Then, by applying  inequalities
\eqref{eq:tight} and \eqref{eq:uniform} and using the definition of
$w_{ij}^{(m)}$, we obtain that
\begin{eqnarray*}
&& - E\bigl[\log\pi(q_{ij}\theta_{ij})|\mathbf{Y}_o,\bolds{\mu
}^{(m)},\mathbf{A}^{(m)},\mathbf{B}
^{(m)}\bigr] \\
&&\qquad  \leq\sum_{q_{ij} = \pm1} \pi\bigl(q_{ij}\theta_{ij}^{(m)}\bigr)
\bigl[ - \log\pi\bigl(\theta_{ij}^{(m)}\bigr)  - \bigl\{1- \pi\bigl(q_{ij}\theta_{ij}^{(m)}\bigr)\bigr\}\bigl\{q_{ij}\bigl(\theta
_{ij}- \theta_{ij}^{(m)}\bigr)\bigr\}\\
&&\qquad\quad\hspace*{200pt} {}+ w_{ij}^{(m)}\bigl\{\bigl(\theta_{ij}- \theta_{ij}^{(m)}\bigr)\bigr\}^2\bigr]\\
&&\qquad  \leq C_m + w_{ij}^{(m)} \bigl\{\bigl(\theta_{ij}- \theta_{ij}^{(m)}\bigr)\bigr\}^2,
\end{eqnarray*}
where $C_m$ is a constant independent of $\bolds{\mu}$, $\mathbf{A}$
and $\mathbf{B}$.
Combining the above results, we see that
$-Q(\bolds{\mu},\mathbf{A},\mathbf{B}|\bolds{\mu}^{(m)},\mathbf
{A}^{(m)},\mathbf{B}^{(m)})$ is up to a constant
majorized by $\sum_{ij} w_{ij}^{(m)} \{(\theta_{ij}-z_{ij}^{(m)})\}^2$,
where $z_{ij}^{(m)}$ equals $x_{ij}^{(m)}$ if $(i,j)\notin\mathcal{N}$,\vspace*{1pt}
and $\theta_{ij}^{(m)}$ if $(i,j)\in\mathcal{N}$. The proof of part (i)
is thus complete. Part (ii) of the theorem follows from the general
result of the MM algorithm. \qed
\end{appendix}

\section*{Acknowledgments}
We would like to thank Editor Michael Stein, an Associate Editor and two
referees for helpful comments. We would also like to thank Lan Zhou for
help in improving the writing of the paper.

\begin{supplement}
\stitle{The MM algorithm for sparse logistic PCA using the tight bound}
\slink[doi]{10.1214/10-AOAS327SUPP}
\slink[url]{http://lib.stat.cmu.edu/aoas/327/supplement.pdf}
\sdatatype{.pdf}
\sdescription{We develop the MM algorithm for sparse logistic PCA using
the tight
majorizing bound. Comparison of the developed algorithm with the MM
algorithm using the uniform bound in terms of computing time is also presented.}
\end{supplement}

\printaddresses

\end{document}